\documentclass[lettersize,journal]{IEEEtran}
\usepackage{generic}
\usepackage{cite}
\usepackage{amsmath,amssymb,amsfonts}
\usepackage{algorithmic}
\usepackage[caption=false,font=normalsize,labelfont=sf,textfont=sf]{subfig}
\usepackage{graphicx}
\usepackage{algorithm,algorithmic}
\usepackage{hyperref}
\usepackage{multirow}
\usepackage{makecell}
\hypersetup{hidelinks=true}
\usepackage{textcomp}
\def\BibTeX{{\rm B\kern-.05em{\sc i\kern-.025em b}\kern-.08em
    T\kern-.1667em\lower.7ex\hbox{E}\kern-.125emX}}
\begin{document}
\title{\textit{DeepCPD}: Deep Learning Based In-Car Child Presence Detection Using WiFi}
\author{Sakila S. Jayaweera, \IEEEmembership{Member, IEEE}, Beibei Wang, \IEEEmembership{Fellow, IEEE}, Wei-Hsiang Wang, \IEEEmembership{Member, IEEE}, and K. J. Ray Liu, \IEEEmembership{Fellow, IEEE}
\thanks{Sakila S. Jayaweera, Beibei Wang, Wei-Hsiang Wang and K.~J.~Ray Liu are with the Origin Research, Rockville, MD 20852 USA (email: sakila.jayaweera@originwirelessai.com; beibei.wang@originwirelessai.com; weihsiang.wang@originwirelessai.com; ray.liu@originwirelessai.com).}
}

\maketitle

\begin{abstract}
Child presence detection (CPD) is a vital technology for vehicles to prevent heat-related fatalities or injuries by detecting the presence of a child left unattended. Regulatory agencies around the world are planning to mandate CPD systems in the near future. However, existing solutions have limitations in terms of accuracy, coverage,  and additional device requirements. While WiFi-based solutions can overcome the limitations, existing approaches struggle to reliably distinguish between adult and child presence, leading to frequent false alarms, and are often sensitive to environmental variations.  
In this paper, we present \textit{DeepCPD}, a novel deep learning framework designed for accurate child presence detection in smart vehicles. \textit{DeepCPD} utilizes an environment-independent feature—the auto-correlation function (ACF) derived from WiFi channel state information (CSI)—to capture human-related signatures while mitigating environmental distortions. A Transformer-based architecture, followed by a multilayer perceptron (MLP), is employed to differentiate adults from children by modeling motion patterns and subtle body size differences. To address the limited availability of in-vehicle child and adult data, we introduce a two-stage learning strategy that significantly enhances model generalization.
Extensive experiments conducted across more than 25 car models and over 500 hours of data collection demonstrate that \textit{DeepCPD} achieves an overall accuracy of 92.86\%, outperforming a CNN baseline by a substantial margin (79.55\%). Additionally, the model attains a 91.45\% detection rate for children while maintaining a low false alarm rate of 6.14\%.

\end{abstract}


\section{Introduction}
\label{sec:introduction}
\IEEEPARstart{W}{ith} the increasing number of child heatstroke deaths reported over the past decades, in-car child presence detection (CPD) systems have become a safety evaluation criterion for next-generation automobiles. The European New Car Assessment Program (Euro NCAP) has introduced a child occupant protection protocol to ensure the protection of children from heart stroke deaths and offers rewards to vehicles that offer CPD solutions satisfying the given protocol.  Toward this end, researchers have been studying the implementation of accurate CPD systems using sensor-based methods, camera-based systems, and radio frequency sensing.

Earlier CPD solutions based on indirect sensing methods, such as pressure, weight, heat, and capacitive sensors\cite{cole2007system, davis2007child, khamil2015babycare, ranjan2013child, capacitive}, cannot distinguish human presence from objects and offer limited coverage.
Later, PIR sensors\cite{hashim2014child, PIR2} were introduced for CPD, detecting children's motion inside the car. However, PIR sensors are sensitive to temperature and environmental impacts and cannot detect children during sleep due to the lack of motion. In contrast to the above methods, vision-based solutions \cite{9522991, muhamad2013car} provide higher accuracy, but they require additional hardware, increasing the deployment cost and energy consumption, and their accuracy relies highly on lighting conditions and cannot work well under obstructions. 
On the other hand, mmWave-based methods\cite{wang2021mmhrv} have gained popularity due to their privacy-preserving capabilities and easy installation. Nevertheless, they face challenges such as limited coverage due to the field of view (FoV) of mmWave antennas and the lack of vehicles equipped with mmWave chipsets.   

Contrary to the above solutions, WiFi sensing provides more extensive coverage, better privacy \cite{8350392, liu_wang_2019, WiFicandomore, Liu2024}, and easy installation with no additional cost, as many new cars already have in-car WiFi. This opens up new in-car sensing applications such as driver monitoring and identification\cite{Qu-Incar, incar-Zeng}, child presence detection\cite{zeng2022wicpd, CPDJayaweera}, and in-car seat occupancy detection. While existing WiFi-based CPD systems provide accurate detection in general use cases, none of them can differentiate between adult and child presence, often triggering unnecessary false alarms when an adult is inside the vehicle. Additionally, severe environmental conditions can further degrade detection performance and increase false alarms, reducing system reliability in challenging situations.  

To tackle the aforementioned challenges, we propose \textit{DeepCPD}—a robust, environment-independent neural network designed to detect child-only presence by effectively distinguishing it from adult and empty-seat scenarios. \textit{DeepCPD} reliably identifies children inside a vehicle regardless of their state (awake or asleep), even in complex real-world environments such as busy parking lots and under severe weather conditions like heavy rain or strong winds. However, designing such a system based on deep learning for different environments presents several challenges.

First, training and evaluating a model that generalizes to any car model requires extensive data collection, but collecting real-world WiFi data, particularly involving children, is challenging. This limitation hampers the model’s ability to generalize to unseen vehicles, often necessitating pre-training for each specific car model. To address this, we collect a comprehensive dataset using diverse antenna configurations and generate synthetic data through data augmentation techniques such as link permutation and link mix, which are introduced based on data pattern analysis to replicate real-world conditions. Furthermore, we introduce a two-stage training strategy that leverages WiFi data collected in residential environments to mitigate the scarcity of motion-specific data.

Second, WiFi sensing deep neural networks that rely on raw channel state information (CSI) inputs often fail to suppress environmental influences, resulting in poor generalization to unseen environments. Achieving robust generalization typically requires a large volume of diverse training data, which takes additional efforts. Moreover, these models are highly dependent on the specific chipset used to collect the CSI, as variations in CSI quality across different hardware can significantly affect performance. In this work, we use the first-order statistics, the Auto-Correlation Function (ACF) of the CSI, as input to the deep neural network. The ACF emphasizes dynamic features related to human motion and breathing patterns while suppressing the effects of static reflections caused by the environment, thereby enabling the network to be both environment and seat-location independent. This approach also reduces dependency on specific hardware, enabling easier deployment in real-world scenarios.

Third, extracting features relevant to child-adult classification remains a significant challenge. While prior works \cite{zeng2022wicpd, CPDJayaweera} have demonstrated the use of ACF for distinguishing presence from non-presence, the features required to differentiate between children and adults are less apparent. However, differences in movement patterns—such as the smaller spatial reflections from children’s bodies and the larger reflections from adults—can be leveraged through spatial analysis. Additionally, children tend to exhibit more abrupt and faster motions, which can be captured through temporal motion patterns. To extract these relevant features, we employ an encoder specifically designed to capture periodic behaviors and cross-subcarrier dependencies. In particular, we adopt the encoder architecture from AutoFormer \cite{AutoFormer}, originally developed for time-series forecasting, due to its ability to model long-range dependencies effectively.

We implement our system using an NXP WiFi chipset with two transmit (Tx) antennas and two receive (Rx) antennas. 
Our dataset includes child data, adult data, and empty data collected over a period of six months for 30 different car models, considering various car models, antenna settings, subject positions, and environmental factors.
Experimental results show that the proposed design achieves an overall accuracy of 92.86\% on unseen test data, significantly outperforming the baseline CNN model by more than 12\%. Moreover, the model attains an average child detection rate of 91.45\% for children aged 0–6 years, with a false alarm rate of 6.14\%. Further analysis reveals that the model maintains a 99\% detection rate for younger children, while performance decreases in scenarios involving awake older children (around 6 years old), likely due to increased similarity in motion patterns between older children and adults.

In summary, our main contributions are as follows:
\begin{enumerate}
\item We propose a novel deep learning framework that leverages an environment-independent ACF representation to capture spatio-temporal patterns for distinguishing child-only presence from adult and empty-seat scenarios.
\item We establish a comprehensive in-car WiFi dataset by collecting data across 30 different car models, covering child presence, adult presence, and empty scenarios.
\item We introduce two synthetic data generation techniques—link permutation and link mix—specifically designed to mimic real-world variations relevant to the task.
\item We demonstrate a two-stage training procedure to address in-car data scarcity, using WiFi data collected in residential environments and validating the transferability of a pre-trained encoder across domains.
\end{enumerate}

The remainder of this article is structured as follows. Section~\ref{literature} presents a review of related work. Section~\ref{feture_extract} delves into the details of environment-independent feature extraction. The design of \textit{DeepCPD} is detailed in Section~\ref{sec:system}. Implementation details and dataset description are provided in Section~\ref{sec:implementaion}, followed by evaluation results in Section~\ref{evaluations}. Finally, Section~\ref{sec:conclusion} concludes the paper.

\section{Related Works}\label{literature}

With the high usage of automobiles and advancements in technology, in-car monitoring applications for smart cars have been developed, focusing on enhancing safety measures and passenger convenience. RF-based sensing has emerged as a promising technology for in-car sensing applications due to its privacy-preserving ability and broader coverage\cite{Qu-Incar, incar-Zeng}. A diverse array of applications using RF technologies like mmWave, UWB, and WiFi have been developed for tasks such as occupant monitoring \cite{ma2020carosense, farsaei2023ieee, wu2022live}, driver authentication and monitoring \cite{wang2021driver, regani2019driver, driver_head_tracking}, activity recognition \cite{wang2019wicar}, and CPD \cite{zeng2022wicpd, CPDJayaweera, novelic, IEE, vayyar}.

Earlier works of CPD relied on physical sensors, e.g., pressure and capacitive sensors, which are mounted to the seats \cite{cole2007system, davis2007child, khamil2015babycare, ranjan2013child, capacitive}. These methods are primarily effective when the child is seated in a baby seat but fail to differentiate between a child and objects of similar weight, and the coverage is only limited to the seating areas, rendering them ineffective if a child moves to the footwell area. While motion sensor-based technologies such as PIR \cite{hashim2014child, PIR2} can identify a child who is awake and usually in motion, a sleeping child without any motion cannot be easily detected. Also, PIR sensors are vulnerable to environmental changes such as temperature.  On the other hand, vision-based solutions \cite{9522991, muhamad2013car} achieve better detection, but require additional hardware, and unobstructed images are needed to achieve accurate detection.  


In contrast, WiFi-based CPD reuses the existing in-car WiFi infrastructure with extensive coverage and user privacy. While existing works demonstrate the potential of detecting adults using WiFi sensing, specifically using breath estimation\cite{TR-breath,respirationWang, zhang2019smars, InCarBreath} and motion sensing\cite{zhang2019widetect, WiSpeed, wiball}, detecting children are more challenging due to their smaller size and subtle movements. 
Previous studies such as WiCPD \cite{zeng2022wicpd} demonstrated good performance, but they did not consider unfavorable environmental conditions that can deteriorate breathing patterns. Our prior work \cite{CPDJayaweera} introduced enhancements to the auto-correlation function (ACF) to bolster the detection performance, but may still fail when the breathing patterns are severely corrupted and restoration may not be feasible. In addition, few prior works have considered distinguishing between the presence of adults and children within a vehicle to reduce false alerts caused by adults. To address all the aforementioned challenges, we propose \textit{DeepCPD}, which can maintain high accuracy even under unfavorable environmental conditions, and is capable of detecting child-only presence, distinguishing it from the previous works.

\section{Domain Independent Feature Extraction}\label{feture_extract}

In this section, we outline the methodology for extracting domain-independent features, covering the channel state modeling, auto-correlation function (ACF) computation, and the efficient design of model inputs.

\subsection{Channel Model}\label{stat_model}

Within the confines of a vehicle, rich multi-path propagation of WiFi signals can undergo reflection, scattering, and diffraction off various surfaces, including human bodies, seats, and the vehicle's structure, before reaching the receiver. This results in the superposition of numerous multi-path components (MPC) at the receiver. Taking this multi-path effect into account, the channel state information (CSI) estimated on a subcarrier on frequency $f$ at time $t$ can be modeled as 

\begin{align}\label{eq:1}
    H(t,f) &= \sum_{l \in \Omega_m} a_l(t)e^{-j2\pi f\tau_l(t)} + n(t,f),  
\end{align}
where $\Omega_m$ denotes the set of multipath components, $a_l(t)$ and $\tau_l(t)$ denote the complex amplitude and the propagation delay of the $l^\mathrm{th}$ multipath component, and $n(t,f)$ represents the additive white Gaussian noise, with power of $\sigma^2(f)$.

Dynamic entities, such as humans, generate time-varying MPCs due to movement, while static objects contribute time-invariant MPCs. Notably, even stationary humans can induce time-varying MPCs, as WiFi signals are sensitive to the subtle movements caused by breathing. Consequently, the estimated CSI,  $H(t,f)$, can be reformulated as
\begin{align}\label{eq:2}
    H(t,f) 
    &= \sum_{m \in \Omega_s} a_m(t)e^{-j2\pi f\tau_m(t)} \nonumber\\ &
    + \sum_{n \in \Omega_d} a_n(t)e^{-j2\pi f\tau_n(t)} + n(t,f) ,
\end{align}
where $\Omega_s$ and $\Omega_d$ denote the time-invariant and time-variant multi-path components, respectively. $a_m(t), a_n(t)$ are the complex amplitudes, $\tau_m, \tau_n$ are the time delay of the $m$ and $n$-th MPC. Assuming that the time-invariant MPCs ($\Omega_s$) remain constant over time, the CSI can be approximated by
\begin{align}\label{eq:3}
    H(t,f) 
    &\approx H_s(f)
    + \sum_{n \in \Omega_d} a_n(t)e^{-j2\pi f\tau_n(t)} + n(t,f) ,
\end{align}
where $H_s(f)$ is the sum of all static MPCs.   
In practical applications, as the phase information usually becomes unreliable due to synchronization offsets \cite{Centermeter_accu, ChenMulti}, we focus on the power response of the CSI  given by

 \begin{align}
        G(t,f) &= |H(t,f)|^2 \nonumber\\
        &= \mu(t,f) + \epsilon(t,f).
\end{align}
where $\mu(t,f)$ denotes the total power of received signal and $\epsilon(t,f)
$ denotes the measurement noise, which can be modeled as additive white Gaussian noise (AWGN) . Also, it can be assume that $\mu(t,f)$ and $\epsilon(t,f)
$ are uncorrelated with each other \cite{zhang2019widetect}. 

\subsection{ACF Statistic Extraction}\label{ACF}

According to electromagnetic wave theory, \cite{zhang2019widetect} demonstrates that the auto-correlation function (ACF) of the CSI power, $\rho(\tau, f)$, is directly correlated with the power of dynamic scatterers, as follows:

\begin{align}
    \rho(\tau, f) &= \frac{\text{cov}[G(t,f), G(t+\tau, f)]}{\text{var}[G(t,f)]} \label{eq:acf}\\
    &= \frac{E_d^2(f)}{E_d^2(f) + \sigma^2(f)} \rho_\mu(\tau,f) + \frac{\sigma^2(f)}{E_d^2(f) + \sigma^2(f)}\delta(\tau)\label{eq:acf2},
\end{align}
where $\tau$ is the time lag, $E_d^2(f)$ and $\sigma^2(f)$ reflects the power of dynamic scatters and the power of measurement noise, respectively. $\rho_\mu(\tau,f)$ is the ACF of $\mu(t,f)$ and $\delta(\tau)$ is the Dirac delta function. 

In a dynamic environment, $E_d^2 > 0$, while in a static environment, $E_d^2 = 0$ as there are no dynamic scatters.  Therefore, $\lim_{\tau \to 0}\rho(\tau,f)$ is an excellent indicator for detecting motion. In practice, the direct measurement of the $\lim_{\tau\to 0}\rho(\tau,f)$ is not feasible due to the limitation of sampling frequency $F_s$. Thus, it can be approximated as $ \lim_{\tau \to 0} \rho(\tau,f) \approx \rho(\tau = 1/F_s, f)$. Thus, motion statistics for the subcarrier $f$, $\psi_f$ can then be defined as,
\begin{align}\label{motion_stat}
   \psi_f = \rho(\tau=1/F_s, f).
\end{align}

The ACF can be further viewed as a periodic signal if there is no motion but breathing \cite{zhang2019smars}. Let us define the channel gain $g(f) \triangleq \frac{E_d^2(f)}{E_d^2(f) + \sigma^2(f)}$. Then for $\tau \neq 0$ we can write the ACF in (\ref{eq:acf2}) as,
\begin{align}
    \rho(\tau, f) = g(f)\rho_\mu(\tau, f).
\end{align}
With the existence of breathing, $\rho_\mu(\tau, f)$ exhibits a periodic pattern with peak values varying across each subcarrier $f$.

For a given WiFi link with $N_s$ subcarriers, we can calculate $N_s$ unique subcarriers with the length of $l$, where $l$ represents the number of time lags. The length $l$ is chosen to effectively capture human breathing patterns.

\subsection{ACF Stability Across Environments}
\begin{figure}[!t]
\centering
\subfloat[]{\includegraphics[trim={0 0 1.3cm 0}, clip,width=1in]{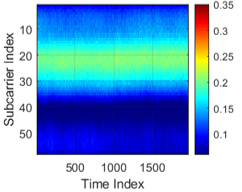}\label{a}}
\hfil
\subfloat[]{\includegraphics[trim={0 0 1.3cm 0}, clip,width=1in]{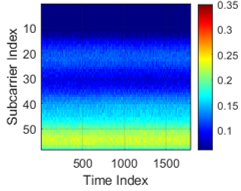}%
\label{b}}
\hfil
\subfloat[]{\includegraphics[trim={0 0 1.3cm 0}, clip,width=1.05in]{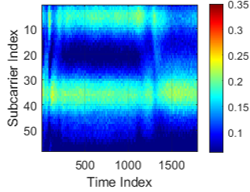}%
\label{c}}

\subfloat[]{\includegraphics[trim={0 0 1.3cm 0}, clip,width=1in]{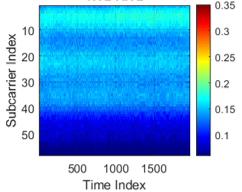}\label{d}}
\hfil
\subfloat[]{\includegraphics[trim={0 0 1.3cm 0}, clip,width=1in]{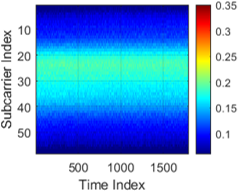}%
\label{e}}
\hfil
\subfloat[]{\includegraphics[trim={0 0 1.3cm 0}, clip,width=1.05in]{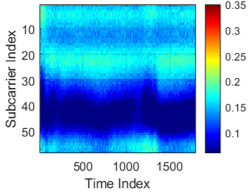}%
\label{f}}
\caption{CSI of (a) a Empty case in Car A, (b) a Child breathing scenario in Car A, (c) a Adult breathing scenario in Car A, (d) a Empty case in Car B, (e) a Child breathing scenario in Car B, and (f) a Adult breathing scenario in Car B.}
\label{CSI_three_case}
\end{figure}

Here, we analyze the CSI and ACF for three scenarios: an empty car, a child breathing and an adult breathing. Fig. \ref{CSI_three_case} shows the CSI for the empty car (a, d), adult presence (c, e), and child presence (b, f) across two car models: Model A (an SUV) and Model B (a sedan). As illustrated, the CSI patterns vary significantly even within the same class, making it challenging to train a generalized classification model. Furthermore, there is no significant difference between CSI data in the empty-car and the child-presence cases, as the child's breathing motion is subtle. Therefore, it is crucial to remove environmental influences and enhance the features to improve the system design.

\begin{figure}[!ht]
\centering
\subfloat[]{\includegraphics[trim={0 0 8mm 0}, clip,width=1.1in]{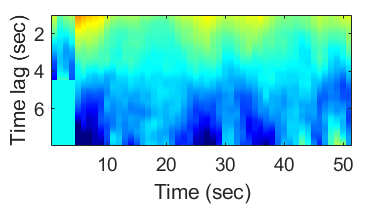}\label{a}}
\hfil
\subfloat[]{\includegraphics[trim={0 0 8mm 0}, clip,width=1.1in]{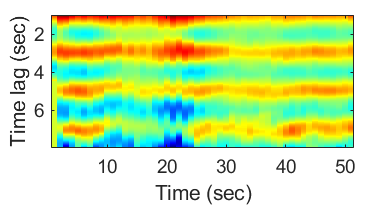}%
\label{b}}
\hfil
\subfloat[]{\includegraphics[trim={0 0 8mm 0}, clip,width=1.1in]{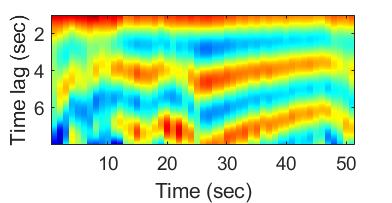}%
\label{c}}

\subfloat[]{\includegraphics[trim={0 0 8mm 0}, clip,width=1.1in]{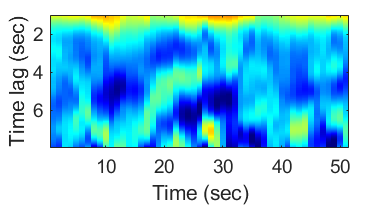}\label{d}}
\hfil
\subfloat[]{\includegraphics[trim={0 0 8mm 0}, clip,width=1.1in]{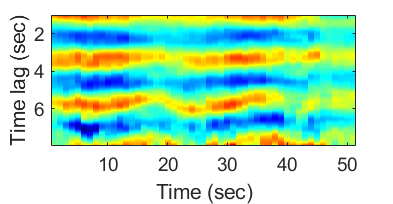}%
\label{e}}
\hfil
\subfloat[]{\includegraphics[trim={0 0 8mm 0}, clip,width=1.1in]{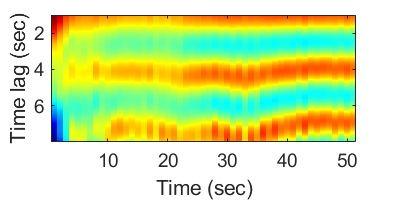}%
\label{f}}
\caption{ACF extracted from most sensitive subcarrier of (a) a Empty case in Car A, (b) a Child breathing scenario in Car A, (c) a Adult breathing scenario in Car A, (d) a Empty case in Car B, (e) a Child breathing scenario in Car B, and (f) a Adult breathing scenario in Car B.}
\label{ACF_three_case}
\end{figure}

Fig.~\ref{ACF_three_case} shows the ACF computed from the most sensitive subcarrier $f$. We select the most sensitive subcarrier based on the motion statistic as explained in Eq.~\ref{motion_stat}. Compared to Fig.~\ref{CSI_three_case}, the ACF in Fig.~\ref{ACF_three_case} highlights more consistent and distinctive activity patterns across different car models, demonstrating its ability to reduce the influence of vehicle-specific characteristics. This suggests that ACF functions as an environment-independent feature by suppressing static components and emphasizing dynamic features, which is valuable for developing a generalized model applicable across different vehicles. While model-based methods can capture some of these distinctions, motion-related ACF patterns can still appear similar across classes, making accurate classification difficult. In contrast, learning-based approaches can more effectively leverage the temporal and spatial characteristics of the ACF, allowing the network to distinguish subtle differences in body size and movement behaviors between children and adults.

\begin{figure*}
    \centering
    \includegraphics[width=4.5in]{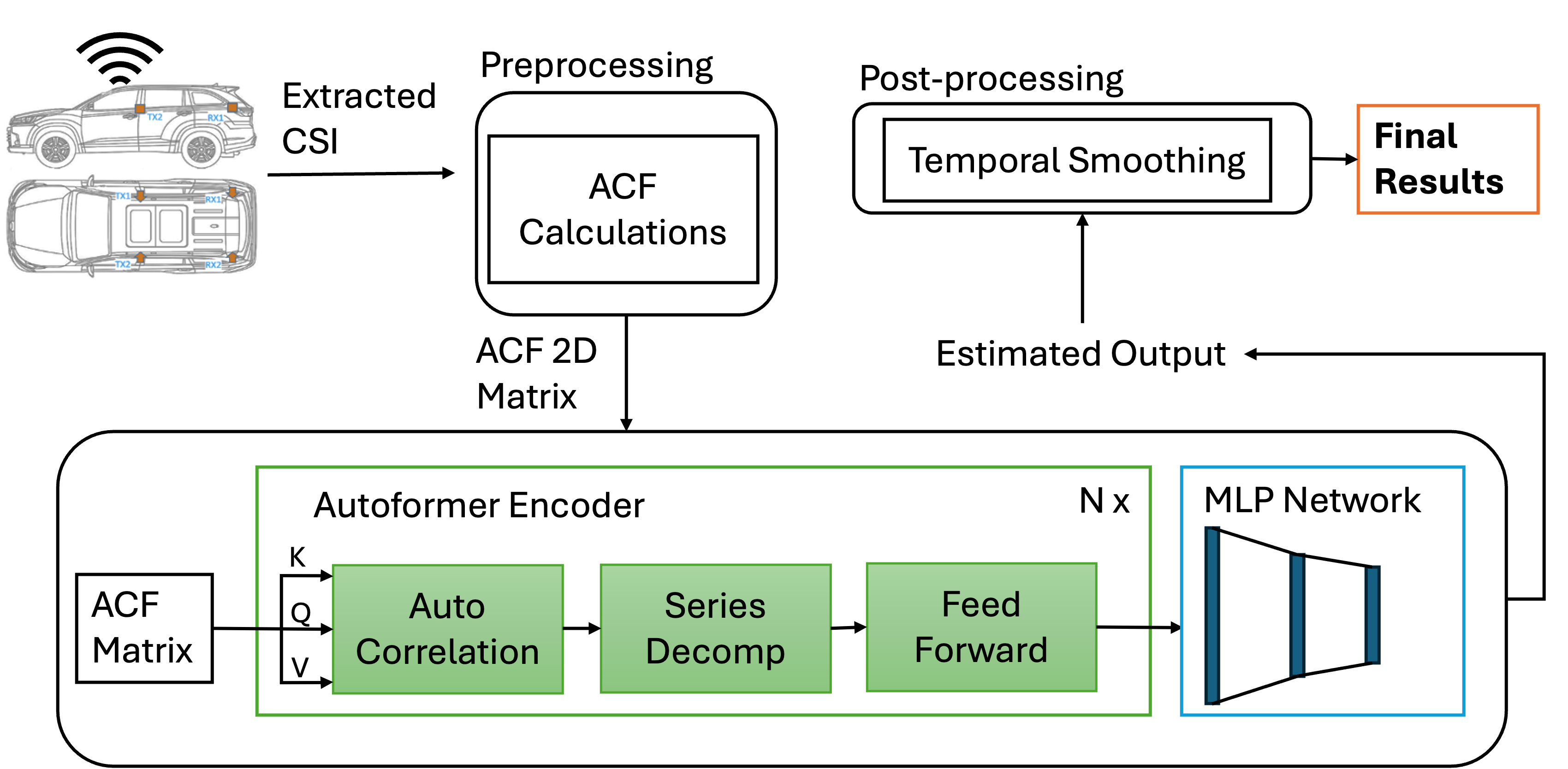}
    \caption{\textit{DeepCPD} System Design}
    \label{fig:system}
\end{figure*}

\section{System Design}\label{sec:system}

As depicted in Fig.~\ref{fig:system}, the \textit{DeepCPD} system is structured into three primary stages. In the first stage, the CSI time series undergoes pre-processing to extract an environment-independent autocorrelation function (ACF) feature matrix. In the second stage, the extracted ACF matrices are fed into the proposed neural network, which is trained to perform three-class classification: empty, child, and adult. In the final stage, post-processing is applied to enhance the classification stability by performing temporal smoothing through window-based fusion over the observation period.

\subsection{Pre-processing}
Motion and respiratory patterns provide the most intuitive and discriminative features for distinguishing child-only presence from other scenarios, such as empty or adult cases. Generally, children’s movements are faster and more abrupt than those of adults, and are associated with smaller body sizes. In addition, when subjects are relatively static, children’s breathing rates are typically higher (approximately 20–30 breaths per minute) compared to adults. Therefore, motion and breathing features extracted from WiFi CSI offer strong cues for differentiating among the three classes: empty, child, and adult.      

However, relying solely on extracted motion statistics and breathing rates may discard other valuable information necessary for robust classification. In addition, noise in the ACF, introduced by environmental factors, can lead to inaccurate estimations. To address these challenges, we propose using the full ACF as the input feature to the network. As discussed in Section~\ref{feture_extract}, the ACF inherently captures information related to both motion and respiratory dynamics.

While previous works~\cite{zhang2019widetect, zeng2022wicpd, CPDJayaweera} have shown that weighted averaging of ACFs across subcarriers can enhance the signal-to-noise ratio (SNR) and improve the accuracy of motion and breathing rate estimation, such aggregation can eliminate important spatial information critical for distinguishing between children and adults. As will be shown in Section~\ref{evaluations}, models trained on averaged ACFs exhibit reduced classification performance.

Since each subcarrier captures reflections from distinct propagation paths and objects, their individual ACF responses contain rich spatial information that can be leveraged for improved child–adult classification. To preserve this diversity, we compute the ACF independently for each subcarrier. These ACFs are then stacked to form a two-dimensional matrix of size $l \times N_s$, where $N_s$ is the total number of subcarriers across all WiFi links, and $l$ is the number of time lags considered when calculating the ACF. The resulting $l \times N_s$ matrix serves as the input to the deep learning model.  


\subsection{Transformer-based Network Design}

The proposed network design consists of two key stages. First, a transformer-based encoder is used to extract spatiotemporal representations from the input data. We adopt the AutoFormer architecture \cite{AutoFormer}, originally designed for time series forecasting, leveraging its inherent decomposition mechanism and ability to model long-range dependencies. In the second stage, an MLP network maps these representations to the final output, enhancing the extraction of discriminative features for classification.

\subsubsection{Positional encoding}
The order of ACF lags contains useful information about signal periodicity and motion patterns, while the order of subcarriers reflects spatial-domain features, such as proximity to the antenna location \cite{proximity}.
Therefore, positional encoding plays a crucial role in preserving these hierarchical structures. Without it, the encoder would treat ACF lags and subcarriers as unordered features, especially when the input sequence is divided into patches along the time (lag) and subcarrier dimensions. To address this, we apply sinusoidal positional encoding across both dimensions, ensuring that each (lag, subcarrier) pair is uniquely encoded to maintain the structural relationships within the data.

\subsubsection{AutoFormer-based encoder}
The Autoformer-based encoder \cite{AutoFormer} consists of three main components: an auto-correlation-based attention block, a series decomposition block, and a feedforward network. 

\textbf{Attention block:} The auto-correlation mechanism replaces the conventional self-attention mechanism used in standard transformer encoders, offering both higher efficiency and improved accuracy for time series analysis \cite{AutoFormer}. This mechanism is particularly effective at extracting periodic patterns related to breathing and motion trends by modeling both subcarrier-local effects and cross-subcarrier dependencies.

Let us assume we have an ACF of length $l$ extracted from $N_s$ subcarriers. In this case, the input X to the encoder layer has the shape $ l \times N_s$. We first project the input X into the query $Q$, key $K$, and value $V$ spaces using learnable linear transformations. We use three separate linear layers to map the input as:
\begin{align}
    Q = X \cdot W_Q, \quad K = X\cdot W_K, \quad V = X \cdot W_V,
\end{align}
where $W_Q, W_K$, and $W_V$ are learnable weight matrices.

For the single-head case, the attention mechanism can be expressed as follows. First, we compute the cross-correlation between Q and K efficiently using FFT operations:
\begin{align} 
R_{(Q, K)}(\tau) = \text{IFFT} (\text{FFT}(Q) \cdot \text{FFT}(K)^{*}),
\end{align}
where $(\cdot)^*$ denotes the complex conjugate of the signal. However, since our input ACF is a real-valued time series, the conjugate simplifies to taking the squared magnitude of the FFT components. 

The resulting cross-correlation scores are used to aggregate the input. For each subarrier, we select the $k$ most correlated lags, $\tau_1, \dots, \tau_k$ and circularly shift ($Roll$) the sequence $V$ by these lags to aggregate the output as below:

\begin{align}
S_{Q,K}(\tau_1), \dots, S_{Q,K}(\tau_k) &= \text{SoftMax}(R_{Q, K}(\tau_1), \dots, \\ & \hspace{3cm} R_{Q, K}(\tau_k))\\
\text{Attention}(Q,K,V) &= \sum_{\tau_i \in top-k} S_{Q,K}(\tau_i) \cdot Roll(V, \tau_i).
\end{align}
where $R_{Q, K}(\tau_i)$ is the autocorrelation of lag $\tau_i$ between series $Q$ and $K$, $Roll(V, \tau_i)$ represents the operation of circular shift with time delay $\tau_i$ and $top-k$ is the subset of time lags with top $k$ correlations. 
Here, we directly aggregated the input based on the most correlated lags, preserving the temporal structure of the sequence.

In the multihead version, the model operates along the subcarrier dimension while preserving both subcarrier relationships and temporal periodicity. Each head i has its own learnable projection matrices, with the query, key, and value dimensions given by $Q^i, K^i, V^i \in \mathcal{R} ^{l\times \frac{N_s}{h} }$, where $i \in {1, \dots, h}$ and $h$ is the number of heads. The multi-head attention output is computed by concatenating the outputs of each head and applying a final linear transformation as,

\begin{align}
\text{Multi-Head Attention} &= W_{out} \cdot \text{Concat}(\text{head}_1, \dots, (\text{head}_h)\\
\text{where } \text{head}_i &= \text{Attention}(Q^i, K^i, V^i).
\end{align}

\textbf{Series decomposition block: }  The decomposition block in the encoder separates the input into trend and seasonal components. This enhances feature quality by isolating periodic patterns from static or slowly varying noise. The seasonal components are analyzed in deeper layers, enabling the model to focus more effectively on finer periodic trends. 

\textbf{Feed forward block:}  The spatiotemporal features extracted by the initial blocks are further processed within the feedforward block to integrate information across subcarriers. This block consists of two 1D convolutional layers, each followed by a Gaussian Error Linear Unit (GELU) activation function to introduce nonlinearity.

\subsubsection{Classification MLP}
An MLP network further processes the encoded feature vector to produce the final classification output. The MLP consists of three fully connected layers, each followed by a ReLU activation function. These layers progressively compress the feature map into a low-dimensional representation mapped to three output values corresponding to the classes: Empty, Adult, and Child. In Section~\ref{sec:implementaion}, we demonstrate the importance of the MLP network for task-specific learning within a two-stage training process.

\subsection{Post-processing}
To mitigate sudden misdetections and false alarms caused by fluctuations between classes, we apply temporal smoothing, which leverages past predictions to reduce spurious spikes. These fluctuations often arise from noise interference in some data samples, leading the model to make abrupt and incorrect class transitions.
As a post-processing step, we apply a moving average filter to the class probabilities using a sliding window. Although this method introduces a small amount of latency proportional to the window size, it offers a favorable trade-off by significantly improving prediction stability and overall system reliability.

\begin{figure*}[!t]
\centering
\subfloat[]{\includegraphics[width=2.35in]{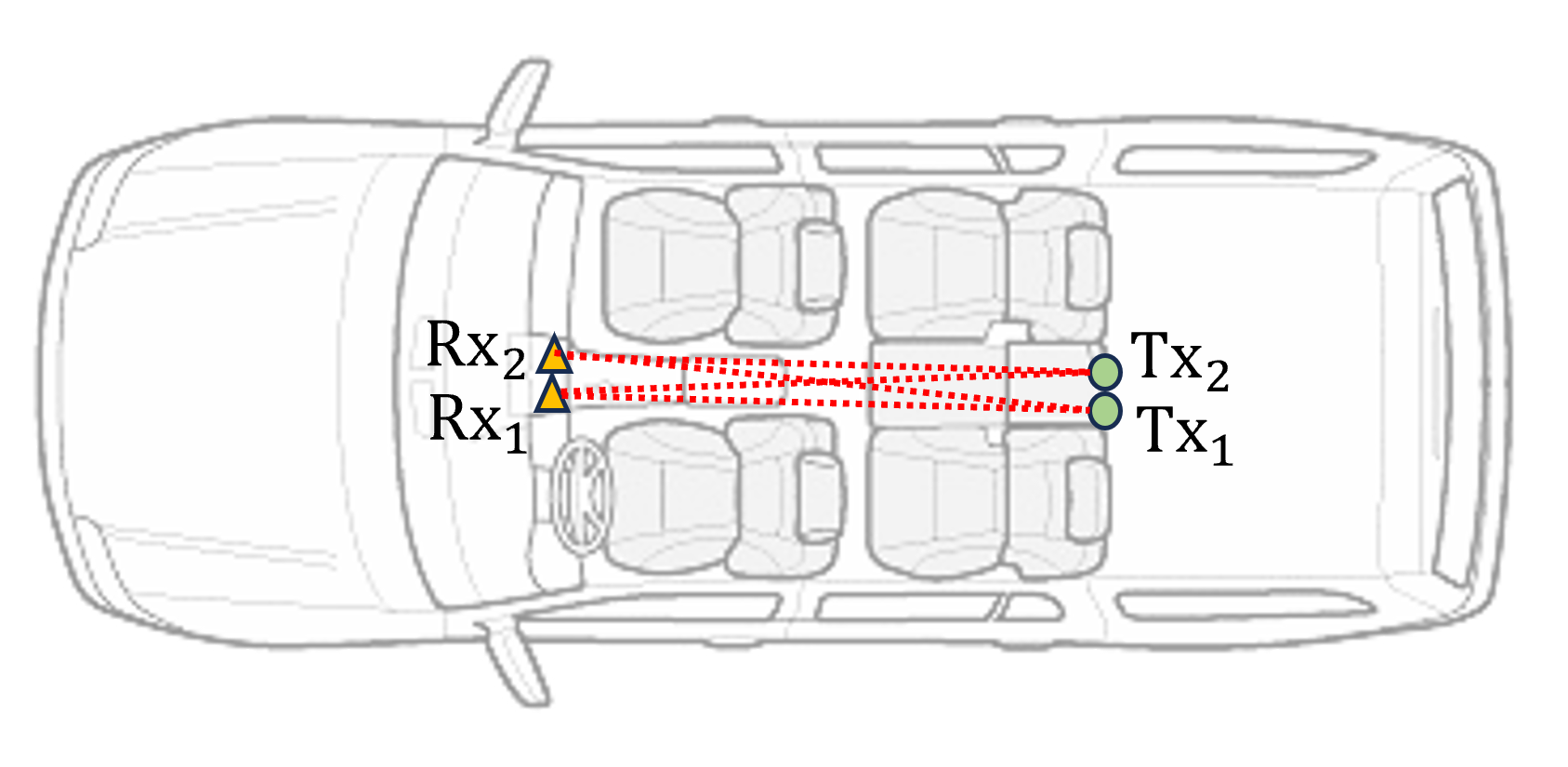}\label{colocated}}
\hfil
\subfloat[]{\includegraphics[width=2.35in]{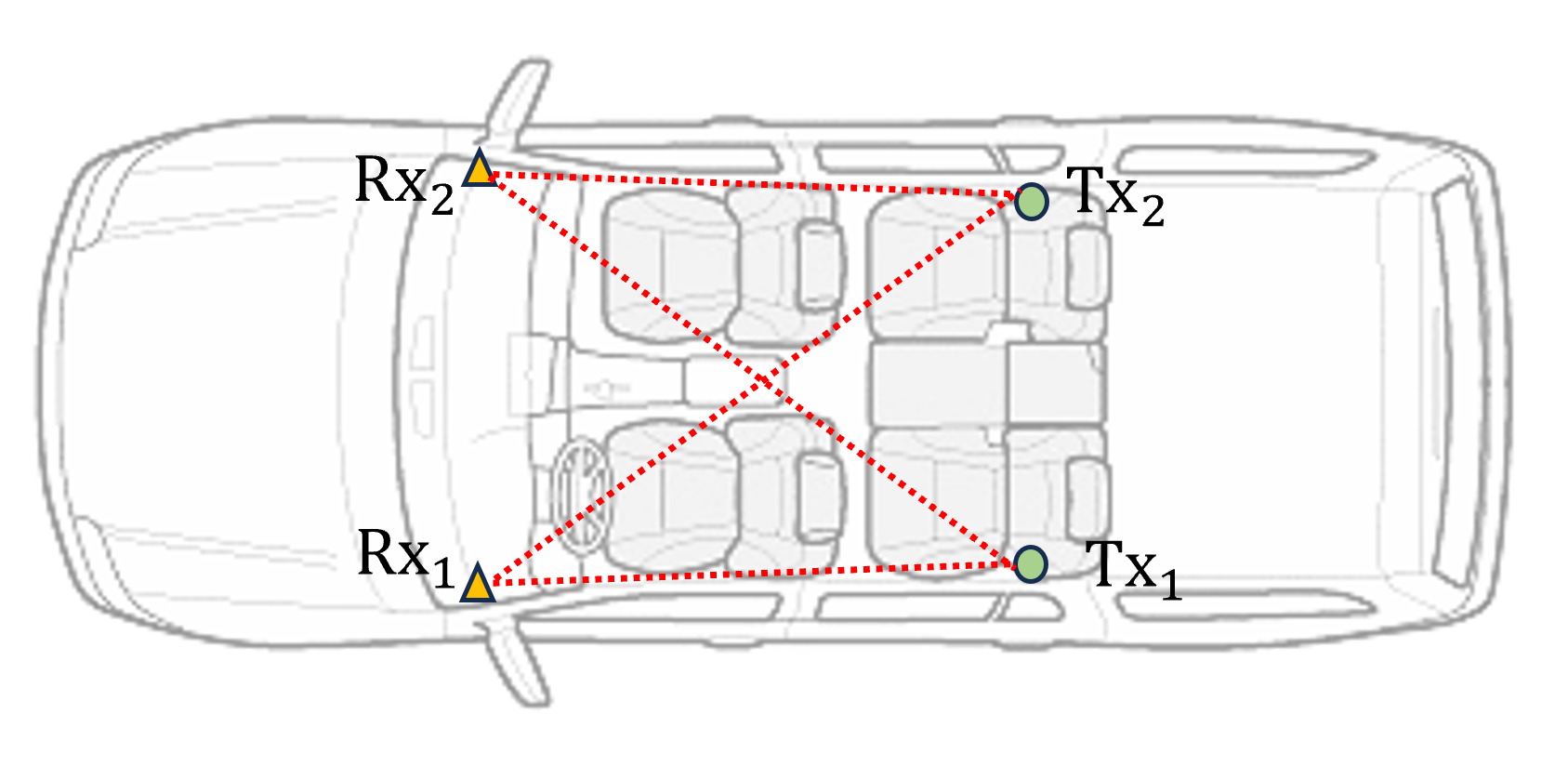}%
\label{distributed}}
\hfil
\subfloat[]{\includegraphics[width=2.35in]{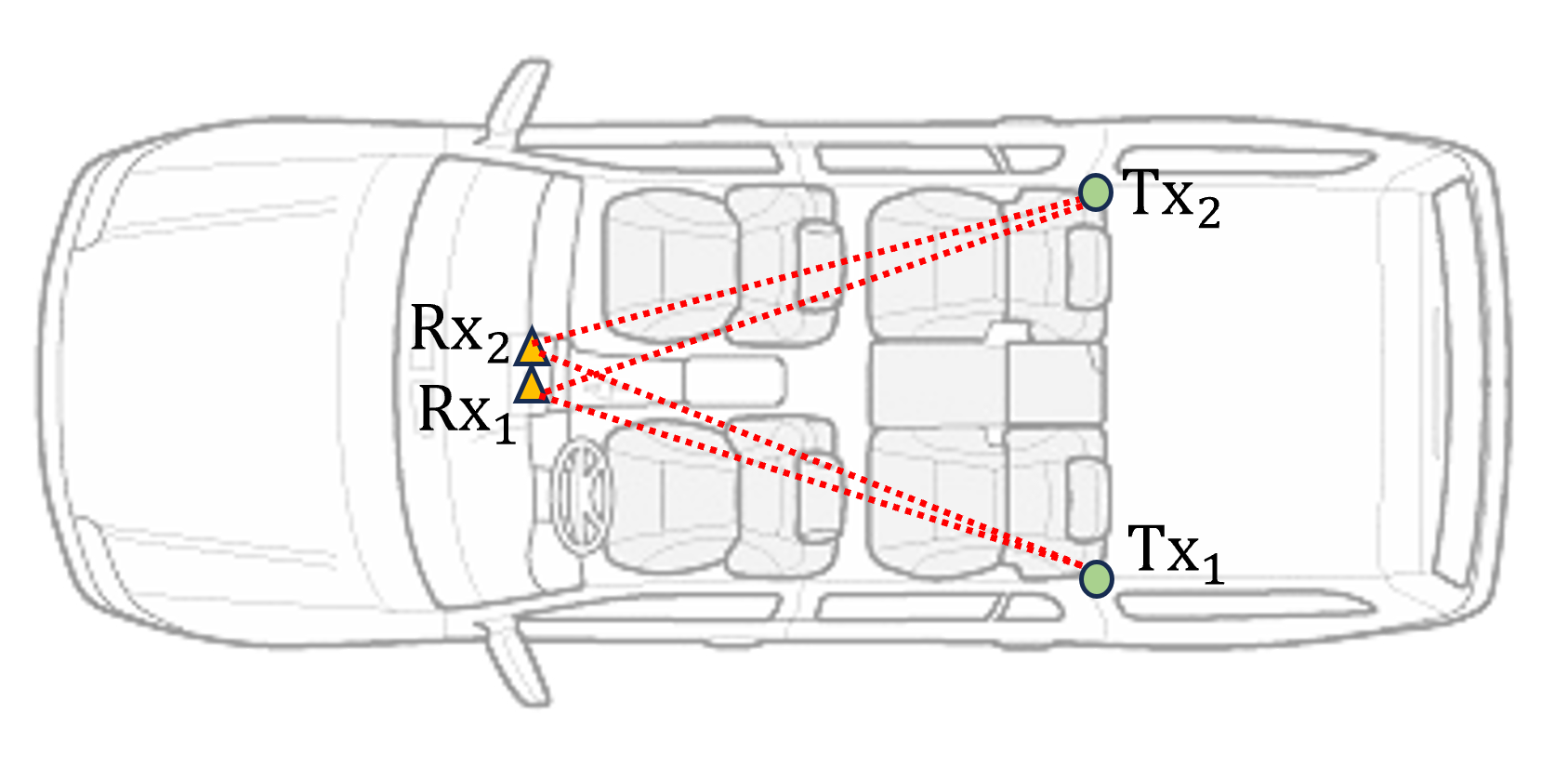}%
\label{mixup1}}
\caption{Different antenna configurations. (a) colocated antenna setup, (b) distributed antenna setup, and (c) hybrid antenna setup.}
\label{antenna_setup}
\end{figure*}

\section{Implementation}\label{sec:implementaion}
We implement our system using a pair of WiFi devices operating at 5 GHz with a bandwidth of 40 MHz. Each device is equipped with two omnidirectional antennas. The estimated CSI at the receiver contains 58 subcarriers with a sample rate of 30 Hz.

\subsection{System Implementation}
\textbf{Antenna configuration:} In our experiments, we collected data under various antenna setups to study their impact on the performance, as detailed below: 
\begin{enumerate}
    \item[C1-] Colocated Setup: Both the transmit (Tx) and receive (Rx) antennas are colocated as shown in Fig. \ref{colocated}. The Rx can be placed in the front, and the Tx can be placed in the back, or vice versa. 
    \item[C2-] Distributed setup: The two Tx and two Rx antennas are placed in the four corners of the car as shown in Fig. \ref{distributed}, where the Rx antennas are placed in the two corners of the front dashboard and Tx antennas are near the two sides of the back seats. 
    \item[C3-] Hybrid setup: The Rx antennas are placed in a colocated manner, and the Tx antennas are placed in a distributed way as in Fig. \ref{mixup1}. The position of Tx and Rx antennas can be interchanged.
\end{enumerate}

\textbf{Data collection:} We conducted extensive data collection, utilizing all three antenna configurations and covering a wide range of conditions. The dataset includes three scenarios:
\begin{enumerate}
    \item Empty case: Data from 30 different car models without occupants.
    \item Child Presence: Data from children under the age of six with their parents’ permission and using a baby doll that can mimic a child’s behavior. Real babies are securely seated in a standard baby seat and positioned across all five seats. The baby doll is tested in both the seating and footwell areas to emulate scenarios where a child might slip from the seat.
    \item Adult Presence: Data from adults over 21 years old using ten different car models. This also includes scenarios with both a child and an adult present. 
\end{enumerate}

Our dataset is built upon the WiCPD\cite{zeng2022wicpd} dataset, which provides a fundamental set of data under normal conditions. The WiCPD dataset includes empty data from 20 different cars, baby doll data from 11 distinct in-car locations, including all seats and footwell positions, and real baby data from five babies secured in car seats. We expand the dataset in WiCPD by incorporating empty and baby data from additional car models and introducing adult presence data.  To study the performance under more challenging conditions, we gathered data in busy parking lots, under adverse weather conditions such as rain/wind, and during challenging interfering motions around the vehicle, such as a person waving a hand next to the car and walking around it. Table \ref{tab:dataset_summary} presents the summary of data in each dataset.

\begin{table*}[]
    \centering
    \caption{Summary of dataset}
    \begin{tabular}{|c|c|c|c|c|p{5cm}|}
    \hline
     & Empty data & Baby Data & Adult data & Antenna configurations & Notes\\
     \hline
    Dataset I \cite{zeng2022wicpd}    & \checkmark &  \checkmark & X & C1 & Data from normal conditions. \\
    \hline
    Dataset II  \cite{CPDJayaweera}   & \checkmark &  \checkmark & X & C1, C2, C3 & Corrupted data from environmental impacts. \\
    \hline
    Dataset III      & \checkmark &  \checkmark & \checkmark & C1, C2, C3 & Data from normal conditions and severely corrupted data due to device noise and environmental effects.  \\
    \hline
    \end{tabular}
    \label{tab:dataset_summary}
\end{table*}

\subsection{Neural Network Implementation}
The deep learning network in \textit{DeepCPD} design is implemented in the PyTorch platform. We train the classifier on an NVIDIA GeForce RTX 4030 GPU for 200 epochs.

\textbf{Two-stage training:} Although we collected a relatively large dataset across 30 different car models, the amount of data from children remains limited. While we can simulate children’s breathing behavior using a baby doll, real child motion data is scarce, and adult data is also somewhat limited. To address the challenge of limited in-car data, we adopt a two-stage training procedure designed to maximize the model’s ability to distinguish subtle motion patterns.

In the first stage, we train the model using labeled data collected from indoor environments, such as houses and apartments, labeled as presence or non-presence, along with in-car data labeled as presence or empty. All adult and child data are grouped into a single class labeled as presence. This training focuses on learning basic motion and breathing features. To maintain consistency with in-vehicle conditions, we exclude walking data from the indoor datasets, as gait patterns are not typically observed inside vehicles. During this stage, the MLP network is adapted for binary classification (empty vs. presence) and trained using binary cross-entropy (BCE) loss with the Adam optimizer. 

In the second stage, we fine-tune the pre-trained encoder using a modified MLP head designed for a three-class classification task. Datasets I, II, and a subset of Dataset III are used for this training. We employ cross-entropy loss with the Adam optimizer to jointly fine-tune the encoder and train the new MLP head. The updated MLP head outputs three probabilities corresponding to the classes: Empty, Adult, and Child. Through this fine-tuning process, the encoder learns to distinguish child-specific features from adult features more effectively.
  
\subsection{Data Augmentation}
To train a generalized neural network, a balanced and sufficiently large dataset is essential. While the two-stage training approach partially addresses the challenges of limited data, we still require a substantial amount of data for each scenario to fine-tune the model effectively. Traditional data augmentation techniques commonly used in computer vision, such as rotation, shifting, and scaling, are not applicable in our case, as preserving temporal information is critical. To overcome this, we introduce two data augmentation techniques specifically designed based on the analysis of data patterns in our application.

\begin{figure*}[ht!]
\centering
\subfloat[]{\includegraphics[trim={0 0 0 0}, clip, width=3in]{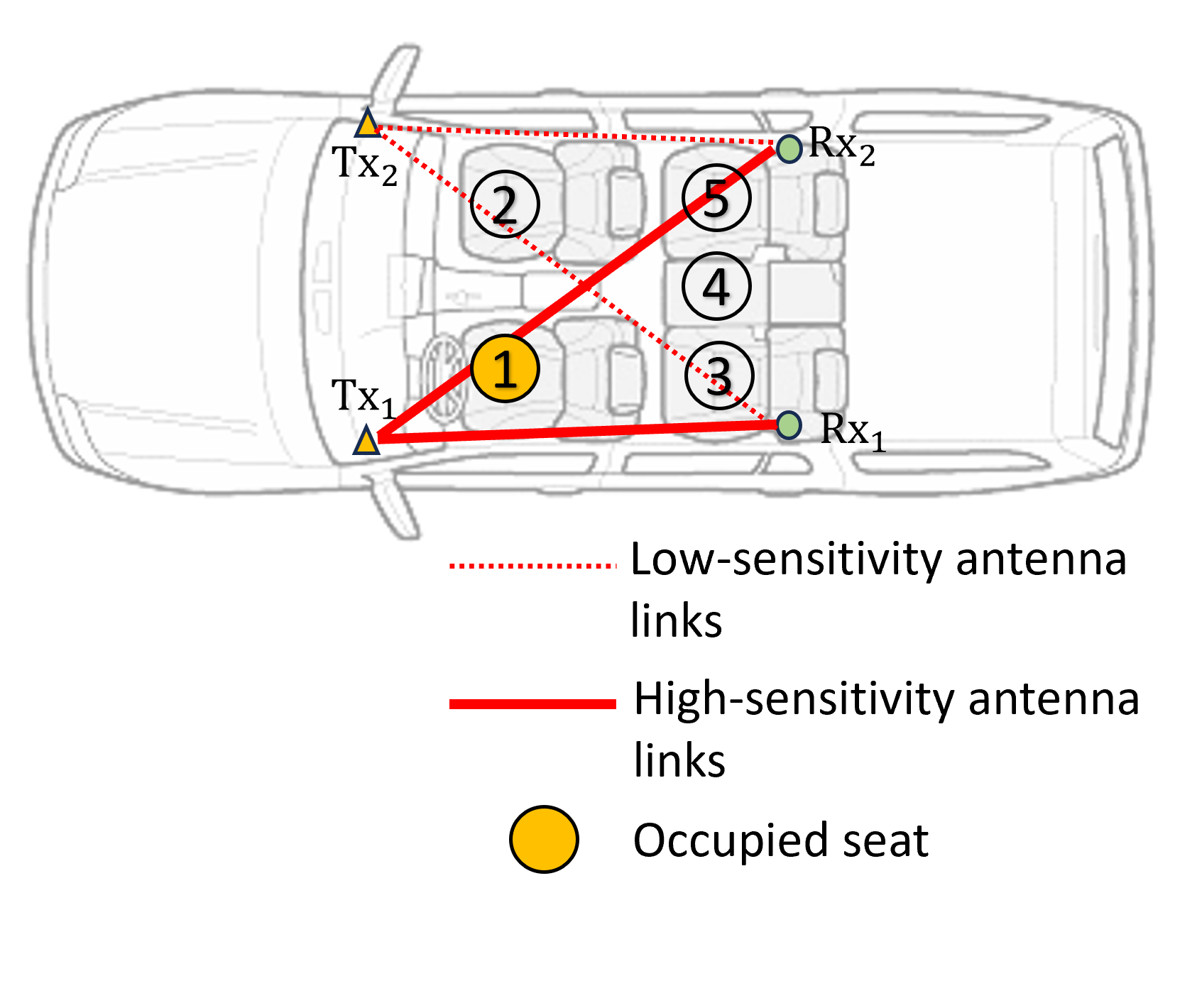}\label{setup}}
\hfil
\subfloat[]{\includegraphics[trim={0 1cm 0 1cm},clip, width=3in, height = 2.8in]{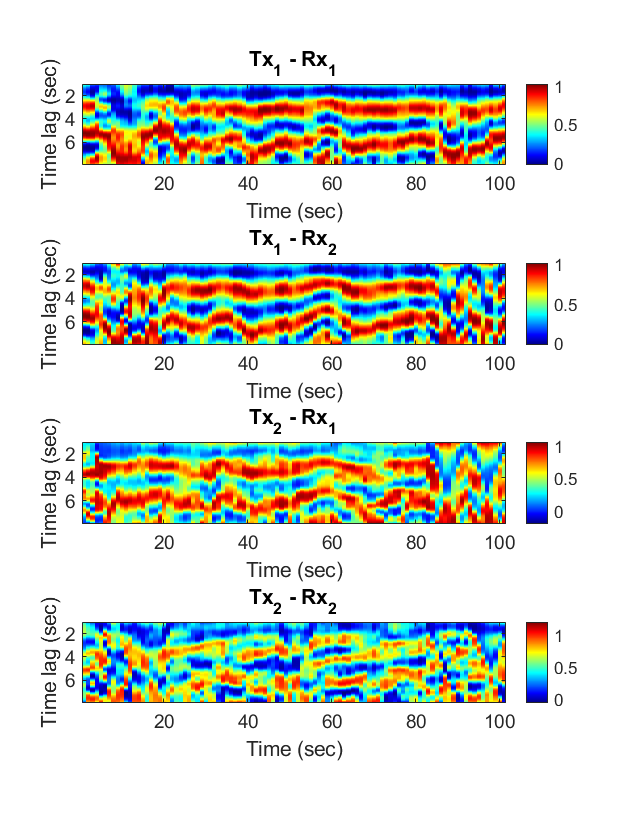}\label{acf}} 
\caption{ACF extracted from each Tx-Rx pair in the breathing scenario when the user occupies seat 1. (a)Antenna configuration and data collection setup, (b) ACF extracted from the highest sensitive subcarrier in each Tx-Rx pair }
\label{fig:acf_links}
\end{figure*}
Here, we analyze the ACF generated from a distributed antenna setup when the user occupies the driver’s seat. With two transmit (Tx) and two receive (Rx) antennas, the system forms four distinct links, each producing a unique ACF. Fig.~\ref{fig:acf_links} shows the ACF extracted from the most sensitive subcarrier by selecting the subcarrier with the highest motion statistic, as explained in Eq.~\ref{motion_stat}, for each link during a breathing scenario when the user is seated in seat 1. As illustrated, the links between Tx1-Rx1 and Tx1-Rx2 exhibit clear periodic patterns, as the user is positioned close to the Tx1 antenna.

\subsubsection{Link permutation}
Building on the observation that links involving the nearest antennas exhibit higher sensitivity, we leverage multi-antenna diversity (in this case, a 2×2 configuration with 2 Tx and 2 Rx antennas) to generate multiple versions of the data. By permuting the autocorrelation functions (ACFs) computed from each transmit-receive antenna pair, we create synthetic samples that simulate variations in user position. This approach effectively augments the dataset, mimicking how signals might behave if the user were sitting in a different position, even though the physical location is not changing.  
This approach helps prevent overfitting to a specific seat location and promotes generalization to unseen antenna configurations (i.e., the same setup with a different antenna ordering).

\subsubsection{Link-mix augmentation}
Inspired by the mix-up data augmentation technique~\cite{zhang2018mixup}, we propose a method that combines samples representing the same activity (child motion) but captured in different environments, positions, or with different Tx-Rx configurations. Specifically, we create a synthetic data sample by merging links from two different samples: high-sensitivity links are taken from one sample, and low-sensitivity links from the other. Sensitivity is measured using the motion statistic feature described in our previous work~\cite{CPDJayaweera, zhang2019widetect}.

Unlike conventional mix-up, where samples from different classes are blended, we only mix data within the same class. This approach generates new, diverse representations of the same activity, improving the network’s ability to generalize to unseen locations (e.g., different car models) and device setups, thereby enhancing domain adaptation.

\section{Evaluations} \label{evaluations}
In this section, we first define the evaluation metrics and then present detailed evaluations, including comparisons with baseline models and performance under different scenarios.

\textbf{Evaluation metric:}
In accordance with the CPD system's requirements, we assess our performance using the following four metrics. 
\begin{itemize}
    \item Detection rate, or True Positive Rate (TPR), which measures the probability of correctly detecting the child's presence.
    \item False alarm or False Positive Rate (FPR), which represents the probability of misidentifying an empty car as a child presence.
    \item Accuracy, which quantifies the overall correctness of the models' predictions, is defined as the ratio of the number of correctly classified data points to the total number of input data.
    \item $F1$ score measures the balance between precision and recall, and is calculated as the harmonic mean of these two metrics as
        \begin{align}
            F1 = 2 \cdot \frac{\text{precision} \cdot \text{recall}}{\text{precision} + \text{recall}},
        \end{align}
    where recall quantifies the likelihood of the system correctly identifying the target, and precision represents the ratio of true positive predictions to the total number of positive predictions made.
\end{itemize}

\subsection{Overall Performance}
\begin{figure}
    \centering
    \includegraphics[width=2.5in]{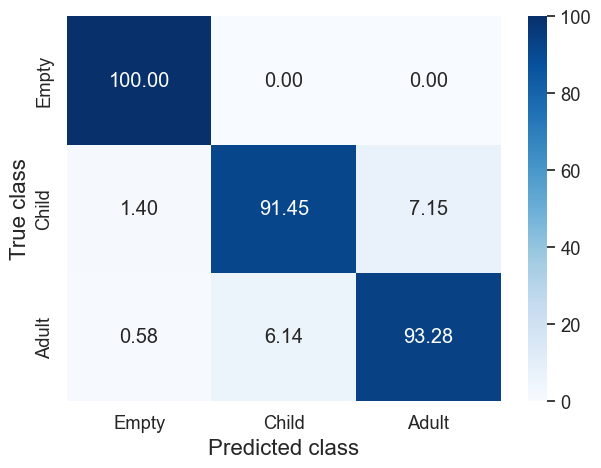}
    \caption{Confusion matrix for child, adult, and empty classification.}
    \label{fig:confusion_DL}
\end{figure}

We utilize datasets I, II, and III, which include data from empty cases, baby presence, and adult presence, to evaluate the overall performance. We categorize instances of child presence with an adult into the adult category to align with our objective of alerting users only when a child is left unattended. We validated the model using $20\%$ of the training dataset and tested it using cars that are not included in the training dataset. The validation and testing accuracy are presented in Table \ref{tab:deep_learning}. The overall test confusion matrix, illustrating the detection performance for each class, is shown in Fig.~\ref{fig:confusion_DL}. As observed, the child detection accuracy is comparatively lower, due to older children’s motion patterns resembling those of adults, leading to misclassification. 

\begin{figure}[ht!]
\centering
\subfloat[]{\includegraphics[width=1.7in]{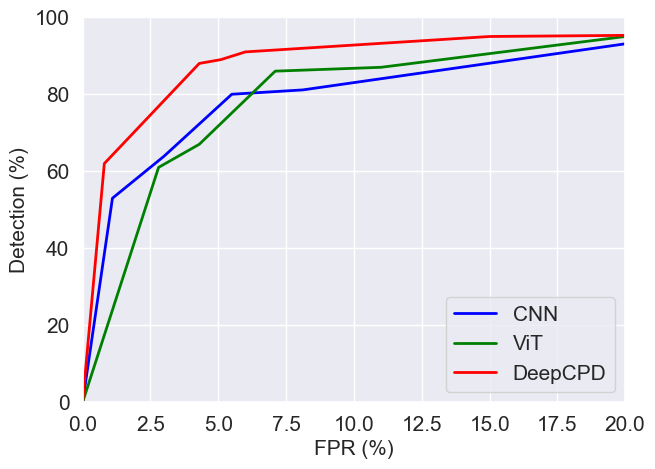}\label{unseen}}
\hfil
\subfloat[]{\includegraphics[width=1.7in]{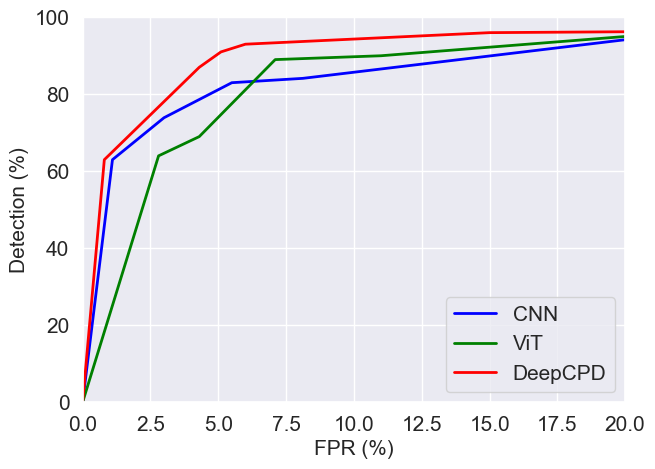}\label{seen}} 
\caption{ROC curves generated from different baseline network architectures. (a)Unseen user in unseen environment, (b) seen user in unseen environment }
\label{fig:roc}
\end{figure}

We compare the validation and testing accuracy of \textit{DeepCPD} against three baseline models, as shown in Table~\ref{tab:baseline_comparison}, and demonstrate that our proposed method consistently outperforms all baselines.
Fig. \ref{fig:roc} compares the ROC curves for the child class using two baseline models, CNN and ViT, alongside our proposed method, \textit{DeepCPD}. As shown, \textit{DeepCPD} outperforms the baselines in both seen and unseen user scenarios. Furthermore, its performance remains consistent even with unseen subjects, demonstrating generalization capability across user variations.

\begin{table}[ht]
    \centering
    \caption{Performance of child and adult classification}
    \begin{tabular}{c|c|c|c|c}
    \hline
    & \multicolumn{2}{c|}{Overall} & \multicolumn{2}{c}{Child class}\\
    \cline{2-5}
    & Accuracy & F1 score &TPR & FPR\\
    \hline
     Validation    &  $97.86\%$ & $96.41\%$& $96.99\%$ & $4.33\%$\\
     \hline
     Test    & $92.86\%$ & $90.11\%$ & $91.45\%$ & $6.14\%$ \\
     \hline
    \end{tabular}
    \label{tab:deep_learning}
\end{table}

\begin{table*}[]
    \centering
    \caption{Comparison with baseline neural network models}
    \begin{tabular}{|c|p{2cm}|c|c|c|c|p{7cm}|}
    \hline
    & & \multicolumn{2}{c|}{Overall Accuracy} & \multicolumn{2}{c|}{Child class} &\\
    \cline{3-6}
    Method & Input & Validation & Test  & TPR & FPR & Comments \\
     \hline
    CNN    & CSI & 96.51\%  & 79.55\% & 68.22\% & 21.73\% & \begin{tabular}[c]{@{}l@{}}
- Hard to generalize across antenna settings \\ 
- Highly dependent on chipset, even with same CSI shape \\
- Requires CSI standardization before processing \\
- Supports only fixed CSI input dimensions
\end{tabular} \\
    \hline
    Shared CNN + MLP & Link level ACF & 88.66\%  & 81.32\% & 78.29\% & 25.81\% & \begin{tabular}[c]{@{}l@{}}
- Works well for separated antenna setting \\ 
- Not generalizable to colocated or hybrid antenna settings\\
- Depends on the quality of each antenna link \\
\end{tabular}\\ 
    \hline
    ViT    & ACF from all subcarriers & 98.10\%  & 82.33\% & 81.22\% & 16.92\% & \begin{tabular}[c]{@{}l@{}}
- Low generalization capability of unseen environments\\ 
- High complex network design\\
\end{tabular}\\
    \hline
    \textbf{\textit{DeepCPD}}   & ACF from all subcarriers & 95.61\%  & \textbf{92.86}\% & \textbf{91.45\%} & \textbf{6.14\%}  & \begin{tabular}[c]{@{}l@{}}
- Better generalization capabilities\\ 
- Can be generalized across different antenna setups\\
\end{tabular}\\
    
    \hline
    \end{tabular}
    \label{tab:baseline_comparison}
\end{table*}

\subsection{Impact of Child Status}
For this comparison, we categorize child presence data as either awake or sleeping by analyzing motion statistics and breathing patterns, as described in our previous work~\cite{CPDJayaweera}. The test accuracies for both cases are summarized in Table~\ref{tab:child_status}. Results show that recognizing awake children remains particularly challenging due to the overlap in motion characteristics with adults. In contrast, the sleeping scenario yields higher accuracy, as the breathing rate ranges of adults and children are more distinct—children typically exhibit higher breathing rates (approximately 20–30 BPM), while adult rates generally range from 12–20 BPM.

\begin{table}[h]
    \centering
    \caption{Performance impact on child status}
    \begin{tabular}{c|c|c}
         & Predicted Child (TPR) & Predicted Adult\\
         \hline
     Awake scenario    & 89.51\%  & 7.20\%\\
     \hline
     Sleeping scenario & 98.39\% & 1.50\% 
    \end{tabular}
    \label{tab:child_status}
\end{table}

\subsection{Impact of Child's Age}
To analyze performance variation with respect to children’s age, we group the data into different age ranges and calculate the detection and false alarm rates for each group. As shown in Table~\ref{tab:child_age}, the results indicate that older children exhibit a comparatively lower detection rate and a higher false alarm rate. This is primarily because their motion patterns often resemble those of adults, making accurate differentiation more challenging.  In contrast, younger children show higher detection accuracy and represent the group most vulnerable, as they are typically unable to exit the vehicle on their own. Nevertheless, our model achieves an 85.82\% detection rate, demonstrating its feasibility for reliable child presence detection. 

\begin{table}[h]
    \centering
    \caption{Performance impact on child age}
    \begin{tabular}{c|c|c}
         & Predicted Child (TPR) & Predicted Adult\\
         \hline
     0-2 years    & 99.51\%  & 0.20\%\\
     \hline
     2-5 years & 93.39\% & 1.50\% \\
     \hline
     6 years &  85.82\% & 10.11\% \\
    \end{tabular}
    \label{tab:child_age}
\end{table}

\subsection{Impact of Different Antenna Setups}
Table~\ref{tab:antenna} illustrates the average testing performance for different antenna setups. While our network can be generalized well across different antenna settings, a colocated antenna setup demonstrates comparatively lower performance. This is because colocated Tx-Rx links capture less diverse spatial information, resulting in less varied ACFs than those obtained from the other two setups. Even though we trained the same model architecture on data from all setups (including colocated antennas), the antenna configuration inherently limits the physical diversity of the input data.
\begin{table}[ht]
    \centering
    \caption{Performance on different antenna setups}
    \begin{tabular}{c|c|c|c|c}
    \hline
    & \multicolumn{2}{c|}{Overall} & \multicolumn{2}{c}{Child class}\\
    \cline{2-5}
    Antenna setup & Accuracy & F1 score &TPR & FPR\\
    \hline
     C1- Colocated     &  $87.68\%$ & $85.17\%$& $82.79\%$ & $12.33\%$\\
     \hline
     C2- Seperated    & $93.00\%$ & $93.11\%$ & $92.56\%$ & $5.11\%$ \\
     \hline
     C3- Hybrid    & $91.66\%$ & $90.99\%$ & $91.75\%$ & $4.68\%$ \\
     \hline
    \end{tabular}
    \label{tab:antenna}
\end{table}

\subsection{Evaluation on Coverage}
A primary benefit of using WiFi for presence detection is its extensive coverage. To validate this assertion, we conducted experiments where a baby doll was positioned in various locations. These included five seating areas, both with and without baby seats, as well as footwell positions. The detection rates summarized in Table~\ref{tab:positions} demonstrate that the system effectively identifies presence across a variety of locations. Notably, these rates are higher than the overall system performance, as the evaluation was conducted using a baby doll that mimics the breathing pattern of an infant under one year old and does not exhibit challenging motion cases.
\begin{table}[ht]
    \centering
     \caption{Detection rate for different positions inside the vehicle}
    \begin{tabular}{c|c}
    \hline
         & Detection rate/ TPR \\
    \hline
    Baby doll with baby seat   & $99.0\%$\\
    Baby doll without baby seat & $99.1\%$\\
    Baby doll in footwell position & $97.4\%$\\
    \hline
    \end{tabular}
   
    \label{tab:positions}
\end{table}

\subsection{Ablation Study} 
In this section, we evaluate the impact of pre-processing steps, input feature types, model architecture components, and training strategies on the final performance of the proposed system. 

\begin{figure}[ht!]
\centering
\subfloat[]{\includegraphics[width=1.7in]{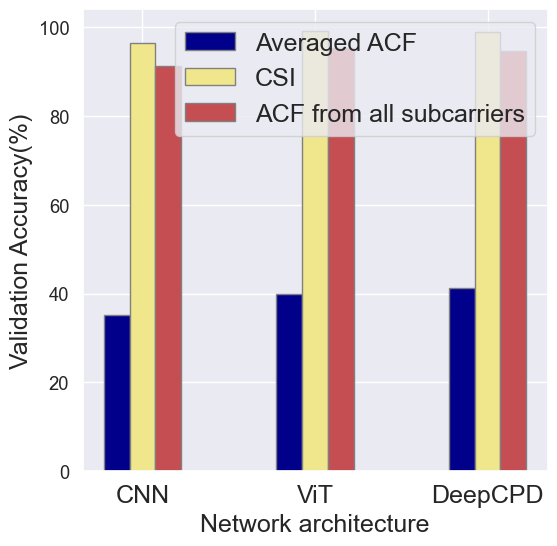}\label{validation}}
\hfil
\subfloat[]{\includegraphics[width=1.7in]{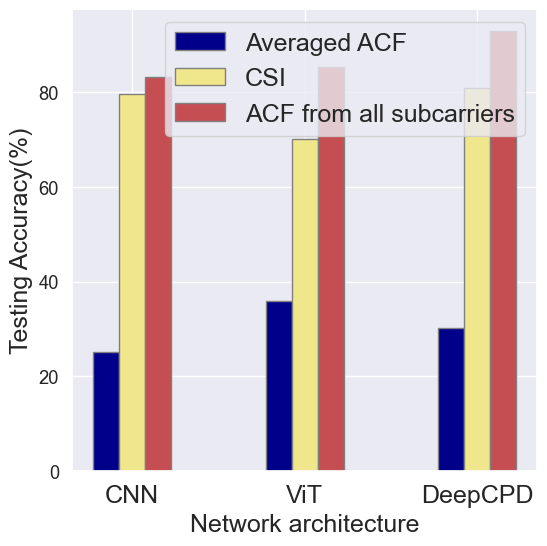}\label{test}} 
\caption{Performance evaluations with different input types. (a) Validation Accuracy, (b) Testing accuracy.}
\label{fig:ablation-input}
\end{figure}

\textbf{Input type: }
To evaluate the effectiveness of using ACF computed from all subcarriers, we compared the performance of the proposed network and baseline models using three input types: raw CSI, averaged ACF, and full subcarrier-wise ACF. The averaged ACF is obtained by combining ACFs across subcarriers using the maximum ratio combining (MRC) approach \cite{zhang2019smars}. Fig.~\ref{fig:ablation-input} shows the validation and testing accuracy for each input type.
Although using raw CSI yields higher validation accuracy, it is not well generalized to new environments, resulting in lower test performance. In addition, although ACF captures motion-related features, averaging it across subcarriers significantly reduces critical information, leading to poor performance in both validation and testing. This loss of spatial detail particularly impacts the ability to distinguish between the presence of children and adults. In contrast, subcarrier-wise ACF consistently achieves the highest test accuracy across all network configurations, demonstrating its robustness to environmental variation and its ability to retain both spatial and temporal features essential for accurate classification. This shows the feasibility of using \textit{DeepCPD} design in real world applications without requiring retraining of each car model/type.

\textbf{Impact of series decomposition block: }
Existing work on AutoFormer~\cite{AutoFormer} demonstrates that the use of a series decomposition block is effective in extracting trend and seasonal components, which are beneficial for time series forecasting. In our work, we adapt the AutoFormer encoder to extract spatio-temporal features relevant to child–adult classification. To assess the impact of the series decomposition block in this context, we evaluate the test accuracy with and without it. Results show an 8\% improvement in accuracy when the block is included, highlighting its importance. 
Specifically, analysis of child breathing data under severe environmental conditions reveals that the block improves the detection rate by over 20\%, demonstrating its ability to enhance subtle breathing patterns, particularly in scenarios where those signals are degraded by environmental noise.


\textbf{Data augmentation: }
We propose two data augmentation strategies that leverage antenna diversity and the data patterns observed in the separated antenna setup to enhance the network’s generalization in unseen environments. Since sleeping children can already be detected with high accuracy without augmentation, we focus here on evaluating the effectiveness of these strategies in the more challenging awake scenario. Table~\ref{tab:data_aug} shows the impact of data augmentation on child-adult classification. As illustrated, the link-mix augmentation method notably improves differentiation between children and adults in the awake state.

\begin{table}[h]
    \centering
    \caption{Performance evaluations with data augmentation}
    \begin{tabular}{p{2cm}|c|c}
         & Predicted Child (TPR) & Predicted Adult\\
         \hline
         \hline
    Without data augmentation   & 72.30\%  & 20.15\%\\
     \hline
     link permutation augmentation  & 77.88\%  & 15.32\%\\
      \hline
     link-mix augmentation    & 82.98\%  & 10.04\%\\
      \hline
     Using both augmentations   & 89.51\%  & 7.20\%\\ 
    \end{tabular}
    \label{tab:data_aug}
\end{table}

\section{Conclusion}\label{sec:conclusion}
This article presents \textit{DeepCPD}, a novel WiFi-based system for child-only presence detection in vehicles using deep learning. The system leverages the environment-independent statistic, the autocorrelation function (ACF), as input to a deep neural network. A transformer-based architecture followed by MLP layers is used to distinguish child presence from adult and empty scenarios.
We implement \textit{DeepCPD} on a commercial WiFi chipset and evaluate its performance using data collected from 25 different vehicle models. Experimental results show that \textit{DeepCPD} achieves an overall accuracy of 92.86\% in unseen environments, outperforming baseline models. Furthermore, the system demonstrates reliability in detecting children with a 91.45\%  detection rate while maintaining a 6.14\%  false alarm rate.

\bibliographystyle{IEEEtran}
\bibliography{references, ref}

\begin{thebibliography}{10}
\providecommand{\url}[1]{#1}
\csname url@samestyle\endcsname
\providecommand{\newblock}{\relax}
\providecommand{\bibinfo}[2]{#2}
\providecommand{\BIBentrySTDinterwordspacing}{\spaceskip=0pt\relax}
\providecommand{\BIBentryALTinterwordstretchfactor}{4}
\providecommand{\BIBentryALTinterwordspacing}{\spaceskip=\fontdimen2\font plus
\BIBentryALTinterwordstretchfactor\fontdimen3\font minus \fontdimen4\font\relax}
\providecommand{\BIBforeignlanguage}[2]{{%
\expandafter\ifx\csname l@#1\endcsname\relax
\typeout{** WARNING: IEEEtran.bst: No hyphenation pattern has been}%
\typeout{** loaded for the language `#1'. Using the pattern for}%
\typeout{** the default language instead.}%
\else
\language=\csname l@#1\endcsname
\fi
#2}}
\providecommand{\BIBdecl}{\relax}
\BIBdecl

\bibitem{cole2007system}
J.~C. Cole, ``System to detect the presence of an unattended child in a vehicle,'' U.S. Patent 7\,170\,401, Jan.~30, 2007.

\bibitem{davis2007child}
L.~Davis, ``Child carseat alert system,'' U.S. Patent 7\,250\,869, Jul.~31, 2007.

\bibitem{khamil2015babycare}
K.~N. Khamil, S.~Rahman, and M.~Gambilok, ``Babycare alert system for prevention of child left in a parked vehicle,'' \emph{ARPN Journal of Engineering and Applied Sciences}, vol.~10, no.~22, pp. 17\,313--17\,319, 2015.

\bibitem{ranjan2013child}
A.~Ranjan and B.~George, ``A child-left-behind warning system based on capacitive sensing principle,'' in \emph{2013 IEEE International Instrumentation and Measurement Technology Conference (I2MTC)}.\hskip 1em plus 0.5em minus 0.4em\relax IEEE, 2013, pp. 702--706.

\bibitem{capacitive}
A.~Rakshitha, A.~George, B.~Babu \emph{et~al.}, ``A child-left behind warning system using capacitive sensing principle,'' \emph{Int. J. Inn. Res. Comput. Commun. Eng}, vol.~5, no.~5, pp. 9526--9531, 2017.

\bibitem{hashim2014child}
N.~Hashim, H.~Basri, A.~Jaafar, M.~Aziz, A.~Salleh, and A.~S. Ja, ``Child in car alarm system using various sensors,'' \emph{ARPN Journal of Engineering and Applied Sciences}, vol.~9, no.~9, pp. 1653--1658, 2014.

\bibitem{PIR2}
P.~Zappi, E.~Farella, and L.~Benini, ``Tracking motion direction and distance with pyroelectric ir sensors,'' \emph{IEEE Sensors Journal}, vol.~10, no.~9, pp. 1486--1494, 2010.

\bibitem{9522991}
J.~Jaworek-Korjakowska, A.~Kostuch, and P.~Skruch, ``{SafeSO}: Interpretable and explainable deep learning approach for seat occupancy classification in vehicle interior,'' in \emph{2021 IEEE/CVF Conference on Computer Vision and Pattern Recognition Workshops (CVPRW)}, 2021, pp. 103--112.

\bibitem{muhamad2013car}
I.~H. Muhamad and F.~R. Rashidi, ``In-car suffocating prevention using image motion detection,'' \emph{Recent Advances in Electrical Engineering Series}, no.~12, 2013.

\bibitem{wang2021mmhrv}
F.~Wang, X.~Zeng, C.~Wu, B.~Wang, and K.~J.~R. Liu, ``{mmHRV}: Contactless heart rate variability monitoring using millimeter-wave radio,'' \emph{IEEE Internet of Things Journal}, vol.~8, no.~22, pp. 16\,623--16\,636, 2021.

\bibitem{8350392}
B.~Wang, Q.~Xu, C.~Chen, F.~Zhang, and K.~J.~R. Liu, ``The promise of radio analytics: A future paradigm of wireless positioning, tracking, and sensing,'' \emph{IEEE Signal Processing Magazine}, vol.~35, no.~3, pp. 59--80, 2018.

\bibitem{liu_wang_2019}
K.~J.~R. Liu and B.~Wang, \emph{Wireless AI: Wireless Sensing, Positioning, IoT, and Communications}.\hskip 1em plus 0.5em minus 0.4em\relax Cambridge University Press, 2019.

\bibitem{WiFicandomore}
C.~Wu, B.~Wang, O.~C. Au, and K.~J.~R. Liu, ``{Wi-Fi} can do more: Toward ubiquitous wireless sensing,'' \emph{IEEE Communications Standards Magazine}, vol.~6, no.~2, pp. 42--49, 2022.

\bibitem{Liu2024}
K.~J.~R. Liu and B.~Wang, ``Statistical principles of time reversal [perspectives],'' \emph{IEEE Signal Processing Magazine}, vol.~41, no.~1, pp. 31--37, 2024.

\bibitem{Qu-Incar}
Q.~Xu, B.~Wang, F.~Zhang, S.~D. Regani, F.~Wang, and K.~J.~R. Liu, ``Wireless ai in smart car: How smart a car can be?'' \emph{IEEE Access}, vol.~8, pp. 55\,091--55\,112, 2020.

\bibitem{incar-Zeng}
X.~Zeng, F.~Wang, B.~Wang, K.~J.~R. Wu, C.and~Liu, and O.~C. Au, ``In-vehicle sensing for smart cars,'' \emph{IEEE Open Journal of Vehicular Technology}, vol.~3, pp. 221--242, 2022.

\bibitem{zeng2022wicpd}
X.~Zeng, B.~Wang, C.~Wu, S.~D. Regani, and K.~J.~R. Liu, ``{WiCPD}: Wireless child presence detection system for smart cars,'' \emph{IEEE Internet of Things Journal}, 2022.

\bibitem{CPDJayaweera}
S.~S. Jayaweera, B.~Wang, X.~Zeng, W.~Wang, and K.~J.~R. Liu, ``Wifi-based robust child presence detection for smart cars,'' in \emph{ICASSP 2023 - 2023 IEEE International Conference on Acoustics, Speech and Signal Processing (ICASSP)}, 2023, pp. 1--5.

\bibitem{AutoFormer}
H.~Wu, J.~Xu, J.~Wang, and M.~Long, ``Autoformer: decomposition transformers with auto-correlation for long-term series forecasting,'' in \emph{Proceedings of the 35th International Conference on Neural Information Processing Systems}, ser. NIPS '21.\hskip 1em plus 0.5em minus 0.4em\relax Red Hook, NY, USA: Curran Associates Inc., 2021.

\bibitem{ma2020carosense}
Y.~Ma, Y.~Zeng, and V.~Jain, ``Carosense: Car occupancy sensing with the ultra-wideband keyless infrastructure,'' \emph{Proceedings of the ACM on Interactive, Mobile, Wearable and Ubiquitous Technologies}, vol.~4, no.~3, pp. 1--28, 2020.

\bibitem{farsaei2023ieee}
A.~Farsaei, B.~Meyer, A.~Sheikh, M.~El~Soussi, P.~Zhang, G.~K. Ramachandra, J.~Govers, and M.~Hijdra, ``An {IEEE} 802.15. 4z-compliant {IR-UWB} radar system for in-cabin monitoring,'' in \emph{2023 IEEE 34th Annual International Symposium on Personal, Indoor and Mobile Radio Communications (PIMRC)}.\hskip 1em plus 0.5em minus 0.4em\relax IEEE, 2023, pp. 1--5.

\bibitem{wu2022live}
Y.~Wu, M.~Zhang, Y.~Shen, and S.~Guo, ``Live body detection method based on millimeter-wave radar in complex car cabin environment,'' in \emph{Proceedings of the 2022 5th International Conference on Telecommunications and Communication Engineering}, 2022, pp. 100--105.

\bibitem{wang2021driver}
F.~Wang, X.~Zeng, C.~Wu, B.~Wang, and K.~J.~R. Liu, ``Driver vital signs monitoring using millimeter wave radio,'' \emph{IEEE Internet of Things Journal}, vol.~9, no.~13, pp. 11\,283--11\,298, 2022.

\bibitem{regani2019driver}
S.~D. Regani, Q.~Xu, B.~Wang, M.~Wu, and K.~J.~R. Liu, ``Driver authentication for smart car using wireless sensing,'' \emph{IEEE Internet of Things Journal}, vol.~7, no.~3, pp. 2235--2246, 2019.

\bibitem{driver_head_tracking}
\BIBentryALTinterwordspacing
X.~Xie, K.~G. Shin, H.~Yousefi, and S.~He, ``Wireless csi-based head tracking in the driver seat,'' in \emph{Proceedings of the 14th International Conference on Emerging Networking EXperiments and Technologies}, ser. CoNEXT '18.\hskip 1em plus 0.5em minus 0.4em\relax New York, NY, USA: Association for Computing Machinery, 2018, p. 112–125. [Online]. Available: \url{https://doi.org/10.1145/3281411.3281414}
\BIBentrySTDinterwordspacing

\bibitem{wang2019wicar}
F.~Wang, J.~Liu, and W.~Gong, ``Wicar: Wifi-based in-car activity recognition with multi-adversarial domain adaptation,'' in \emph{Proceedings of the International Symposium on Quality of Service}, 2019, pp. 1--10.

\bibitem{novelic}
\BIBentryALTinterwordspacing
{Novelic Inc.} Automotive in-cabin monitoring. [Online]. Available: \url{https://www.novelic.com/acam-automotive-in-cabin-monitoring-radar/}
\BIBentrySTDinterwordspacing

\bibitem{IEE}
\BIBentryALTinterwordspacing
{IEE-sensing Inc.} Safety and comfort in-vehicle sensing solutions. [Online]. Available: \url{https://iee-sensing.com/automotive/safety-and-comfort/}
\BIBentrySTDinterwordspacing

\bibitem{vayyar}
\BIBentryALTinterwordspacing
{Vayyar Inc}. Detect everything. protect everyone. [Online]. Available: \url{https://vayyar.com/auto/}
\BIBentrySTDinterwordspacing

\bibitem{TR-breath}
C.~Chen, Y.~Han, Y.~Chen, H.~Q. Lai, F.~Zhang, B.~Wang, and K.~J.~R. Liu, ``{TR-BREATH}: Time-reversal breathing rate estimation and detection,'' \emph{IEEE Transactions on Biomedical Engineering}, vol.~65, no.~3, pp. 489--501, 2018.

\bibitem{respirationWang}
F.~Wang, F.~Zhang, C.~Wu, B.~Wang, and K.~J.~R. Liu, ``Respiration tracking for people counting and recognition,'' \emph{IEEE Internet of Things Journal}, vol.~7, no.~6, pp. 5233--5245, 2020.

\bibitem{zhang2019smars}
F.~Zhang, C.~Wu, B.~Wang, M.~Wu, D.~Bugos, H.~Zhang, and K.~J.~R. Liu, ``{SMARS}: Sleep monitoring via ambient radio signals,'' \emph{IEEE Transactions on Mobile Computing}, vol.~20, no.~1, pp. 217--231, 2019.

\bibitem{InCarBreath}
M.~Hussain, A.~Akbilek, F.~Pfeiffer, and B.~Napholz, ``In-vehicle breathing rate monitoring based on wifi signals,'' in \emph{2020 50th European Microwave Conference (EuMC)}, 2021, pp. 292--295.

\bibitem{zhang2019widetect}
F.~Zhang, C.~Wu, B.~Wang, H.~Q. Lai, Y.~Han, and K.~J.~R. Liu, ``{WiDetect}: Robust motion detection with a statistical electromagnetic model,'' \emph{Proceedings of the ACM on Interactive, Mobile, Wearable and Ubiquitous Technologies}, vol.~3, no.~3, pp. 1--24, 2019.

\bibitem{WiSpeed}
F.~Zhang, C.~Chen, B.~Wang, and K.~J.~R. Liu, ``{WiSpeed}: A statistical electromagnetic approach for device-free indoor speed estimation,'' \emph{IEEE Internet of Things Journal}, vol.~5, no.~3, pp. 2163--2177, 2018.

\bibitem{wiball}
F.~Zhang, C.~Chen, B.~Wang, H.-Q. Lai, Y.~Han, and K.~J.~R. Liu, ``{WiBall}: A time-reversal focusing ball method for decimeter-accuracy indoor tracking,'' \emph{IEEE Internet of Things Journal}, vol.~5, no.~5, pp. 4031--4041, 2018.

\bibitem{Centermeter_accu}
C.~Chen, Y.~Chen, Y.~Han, H.-Q. Lai, and K.~J.~R. Liu, ``Achieving centimeter-accuracy indoor localization on wifi platforms: A frequency hopping approach,'' \emph{IEEE Internet of Things Journal}, vol.~4, no.~1, pp. 111--121, 2017.

\bibitem{ChenMulti}
C.~Chen, Y.~Chen, Y.~Han, H.-Q. Lai, F.~Zhang, and K.~J.~R. Liu, ``Achieving centimeter-accuracy indoor localization on wifi platforms: A multi-antenna approach,'' \emph{IEEE Internet of Things Journal}, vol.~4, no.~1, pp. 122--134, 2017.

\bibitem{proximity}
Y.~Hu, M.~Z. Ozturk, B.~Wang, C.~Wu, F.~Zhang, and K.~J.~R. Liu, ``Robust passive proximity detection using wi-fi,'' \emph{IEEE Internet of Things Journal}, vol.~10, no.~7, pp. 6221--6234, 2023.

\bibitem{zhang2018mixup}
\BIBentryALTinterwordspacing
H.~Zhang, M.~Cisse, Y.~N. Dauphin, and D.~Lopez-Paz, ``mixup: Beyond empirical risk minimization,'' in \emph{International Conference on Learning Representations}, 2018. [Online]. Available: \url{https://openreview.net/forum?id=r1Ddp1-Rb}
\BIBentrySTDinterwordspacing

\end{thebibliography}

\end{document}